\documentclass[longbibliography, aps, pra, amsmath, amssymb, amsfonts, twocolumn,superscriptaddress, footinbib]{revtex4-2}
\usepackage[german,english]{babel}
\usepackage[utf8]{inputenc}

\usepackage{graphicx}
\usepackage{dcolumn}
\usepackage{bm}
\usepackage{physics}

\usepackage{dcolumn}
\usepackage{subfigure}
\usepackage{float}

\usepackage{tabularx}
\usepackage[normalem]{ulem}
\usepackage[usenames, dvipsnames]{color}

\usepackage[linesnumbered, ruled]{algorithm2e}

\usepackage{colortbl}
\usepackage{diagbox}
\usepackage[version=4]{mhchem}

\usepackage{theorem}
\newtheorem{theorem}{Theorem}
\newtheorem{definition}[theorem]{Definition}

\usepackage{hyperref}
\hypersetup{                    
    colorlinks,
    citecolor=blue,
    filecolor=blue,
    linkcolor=blue,
    urlcolor=blue
}

\newcommand{\bmth}{\bm{\theta}}
\newcommand{\red}[1]{\textcolor{red}{#1}}
\newcommand{\wt}[1]{\widetilde{#1}}
\newcommand{\mr}[1]{\mathrm{#1}}

\begin{document}
\title{Accelerated variational quantum eigensolver with joint Bell measurement}

\author{Chenfeng Cao}
\email{chenfeng.cao@connect.ust.hk}
\affiliation{Department of Physics, The Hong Kong University of Science and Technology, \\Clear Water Bay, Kowloon, Hong Kong, China}
\affiliation{QunaSys Inc., Aqua Hakusan Building 9F, 1-13-7 Hakusan, Bunkyo, Tokyo 113-0001, Japan}

\author{Hiroshi Yano}
\affiliation{Department of Applied Physics and Physico-Informatics, Keio University, Hiyoshi 3-14-1, Kohoku, Yokohama 223-8522, Japan}
\affiliation{QunaSys Inc., Aqua Hakusan Building 9F, 1-13-7 Hakusan, Bunkyo, Tokyo 113-0001, Japan}

\author{Yuya O. Nakagawa}
\email{nakagawa@qunasys.com}
\affiliation{QunaSys Inc., Aqua Hakusan Building 9F, 1-13-7 Hakusan, Bunkyo, Tokyo 113-0001, Japan}

\begin{abstract}
The variational quantum eigensolver (VQE) stands as a prominent quantum-classical hybrid algorithm for near-term quantum computers to obtain the ground states of molecular Hamiltonians in quantum chemistry. However, due to the non-commutativity of the Pauli operators in the Hamiltonian, the number of measurements required on quantum computers increases significantly as the system size grows, which may hinder practical applications of VQE. In this work, we present a protocol termed \textit{joint Bell measurement VQE} (JBM-VQE) to reduce the number of measurements and speed up the VQE algorithm. Our method employs joint Bell measurements, enabling the simultaneous measurement of the absolute values of all expectation values of Pauli operators present in the Hamiltonian. In the course of the optimization, JBM-VQE estimates the absolute values of the expectation values of the Pauli operators for each iteration by the joint Bell measurement, while the signs of them are measured less frequently by the conventional method to measure the expectation values. Our approach is based on the empirical observation that the signs do not often change during optimization. We illustrate the speed-up of JBM-VQE compared to conventional VQE by numerical simulations for finding the ground states of molecular Hamiltonians of small molecules, and the speed-up of JBM-VQE at the early stage of the optimization becomes increasingly pronounced in larger systems. Our approach based on the joint Bell measurement is not limited to VQE and can be utilized in various quantum algorithms whose cost functions are expectation values of many Pauli operators.
\end{abstract}

\date{\today}
\maketitle

\section{Introduction}
\begin{figure}[t]
	\centering
	\includegraphics[width=7.5cm]{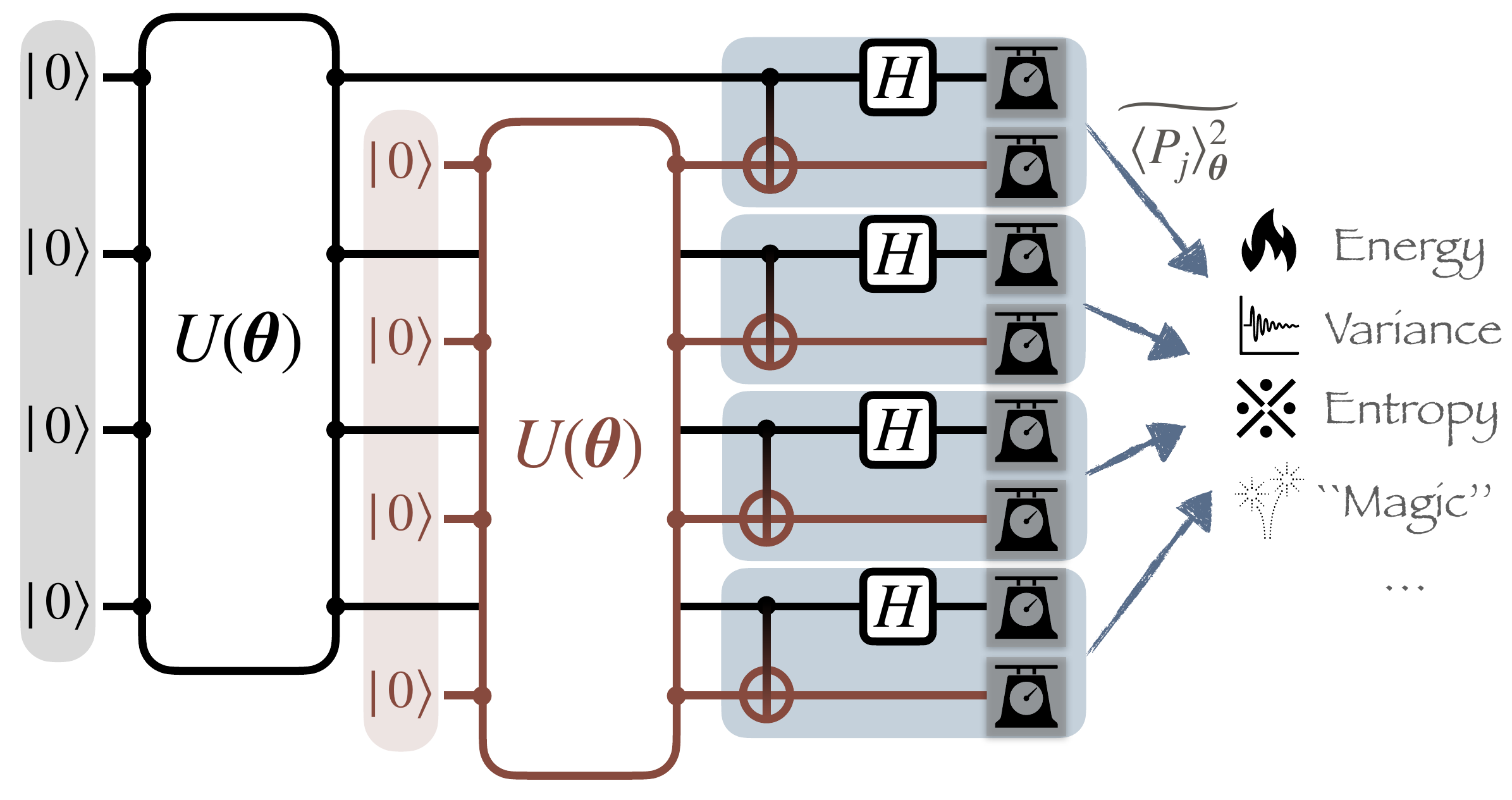}
	\caption{Schematic picture of the joint Bell measurement and JBM-VQE.
    For an $n$-qubit quantum state $\ket{\psi(\bmth)} = U(\bmth)\ket{0}$,
    the $2n$-qubit quantum states $\ket{\psi(\bmth)} \otimes \ket{\psi(\bmth)}$ is prepared and measured in the Bell basis.
    The measurement results can give the estimates of all expectation values $\ev{P_j}_{\bmth}^2 = (\ev{P_j}{\psi(\bmth)})^2$ in the original $n$-qubit systems, and the state information (energy in JBM-VQE) is estimated by these estimates as well as the pre-given signs of the expectation values.}
	\label{fig:schematic}
\end{figure}

Noisy intermediate-scale quantum (NISQ) devices~\cite{Preskill2018} have attracted considerable interest as they hold the potential to solve certain computational tasks faster than classical computers~\cite{arute2019quantum, zhong2020quantum, madsen2022quantum}. 
These devices typically have a relatively small number of qubits (usually between 50 and a few hundred) and are subject to hardware noise, so the computational results obtained from them may not be completely reliable.
Despite these limitations, NISQ devices are expected to be capable of performing calculations that are beyond the capabilities of classical computers, which makes them exciting tools in the near future.
The research community has made substantial progress in developing NISQ-friendly variational quantum optimization algorithms for a variety of applications, including quantum machine learning~\cite{Mitarai2018}, fidelity estimation~\cite{Cerezo2020, Chen2022B}, quantum error-correcting code discovery~\cite{Cao2022}.
The variational quantum eigensolver (VQE) is considered a flagship algorithm within this field, utilizing parameterized quantum circuits and the variational principle to prepare the ground or excited states of quantum many-body systems~\cite{Peruzzo2014, Tilly2022}.

VQE has already been experimentally realized on actual quantum hardware to solve small-sized problems in quantum chemistry and material calculation~\cite{Peruzzo2014, Kandala2017}.
However, the scalability of VQE to larger system sizes remains a challenge.
One of the main reasons, especially in the application to quantum chemistry, is the large number of measurements required during the optimization process.
For the molecular Hamiltonians in quantum chemistry, there are usually $\order{n^4}$ Pauli operators for an $n$-qubit system, and estimation of their expectation values is required in each iteration of VQE.
For example, it was estimated that a single evaluation of the expectation value of the Hamiltonian for analysing the combustion energies of some organic molecules requires $\sim 10^9$ measurement shots and may take as long as several days~\cite{Gonthier2022}. %
To tackle the problem of scalability, various methods have been proposed to reduce the number of measurements in the optimization of VQE.
One possible strategy is to divide the $\order{n^4}$ Pauli operators in the Hamiltonian into the groups of simultaneously-measurable operators, and there are methods that realize $\mathcal{O}(n^3)$ groups~\cite{gokhale2020ON3,zhao2020} and even $\mathcal{O}(n^2)$ groups~\cite{Bonet-Monroig2020, Inoue2023}.
The reduction of the number of groups can result in the reduction of the number of measurements to estimate their expectation values, which leads to the alleviation of the scalability problem of VQE.
Nonetheless, it is still highly demanded to develop a method to reduce the number of measurements in the whole optimization process of VQE. 

Although it is impossible to perform the projective measurement simultaneously on the non-commuting Pauli operators to estimate their expectation values, the \textit{absolute values} of the expectation values can be estimated  simultaneously by using the so-called joint Bell measurement in a doubled system consisting of $2n$ qubits (see Sec.~\ref{subsec:JBM} for detailed explanations).
This is because for any $n$-qubit Pauli operators $P_1, P_2, \dots, P_M$ acting on the original system of $n$ qubit, the operators $P_1 \otimes P_1, P_2 \otimes P_2, \dots, P_M \otimes P_M$ acting on the doubled system of $2n$ qubits commute with each other.
Then the expectation values of the doubled state $(\bra{\psi}\otimes\bra{\psi})(P_j \otimes P_j)(\ket{\psi}\otimes\ket{\psi}) = (\ev{P_j}{\psi})^2$ can be estimated simultaneously for all $j=1,\cdots,M$, where $\ket{\psi}$ is a state in the original $n$-qubit system. The joint Bell measurement is pivotal in numerous domains of quantum information science, including quantum computing, quantum communications, and quantum metrology. This technique was utilized to show an exponential advantage of quantum computers over classical ones in predicting properties of physical systems~\cite{Huang2022, huang2021}, estimate the reduced density matrices~\cite{Jiang2020optimalfermionto}, and learn the nonstabilizerness of a quantum state~\cite{Haug2023Scalable}. In photonic systems, the joint Bell measurement has facilitated the implementation of an optimal orienteering protocol~\cite{Tang2020} as well as breaking the classical limit in quantum target identification~\cite{Xu2021}.

It is noteworthy to point out that the reduction of the number of measurements by a constant factor was achieved in Ref.~\cite{Hamamura2020} through the use of Bell measurement.
While the classical shadow technique~\cite{Huang2020} also seeks to predict the expectation values of numerous operators simultaneously, the measurement overhead in its protocol with random Pauli measurements scales exponentially with the operator's locality. This makes its application to quantum chemistry Hamiltonians with non-local Pauli operators less straightforward (see also Ref.~\cite{Zhao2021}).

In this study, we introduce a method to reduce measurement overhead in VQE by employing joint Bell measurements (JBM), referred to as JBM-VQE (a schematic illustration is provided in Fig.~\ref{fig:schematic}). Our approach takes advantage of the correlation between Pauli expectation values across successive iterations of VQE. More concretely, we observe that the signs of expectation values of the Pauli operators in the Hamiltonian change infrequently during the optimization of VQE while their absolute values change frequently. 
In JBM-VQE, one measures the absolute values by the joint Bell measurement, which requires only one circuit to measure, during every iteration. 
On the other hand, the signs of expectation values of the Pauli operators are measured once in the fixed number of iterations by the conventional measurement method (using naively $\order{n^4}$, at least $\order{n^2}$, distinct circuits).
The expectation value of the Hamiltonian is subsequently constructed by combining the estimated absolute values and signs with assuming that the latest estimation of the signs is valid for the following iterations.
We can expect a reduction in the measurement cost during the optimization by using this protocol.
To exemplify this, we numerically compare JBM-VQE to the conventional VQE for molecular Hamiltonians of small molecules under the reasonable condition that the statistical fluctuations of the energy expectation values in both methods are almost the same.
JBM-VQE requires fewer shots to approach the vicinity of the exact ground state compared to conventional VQE, with this trend becoming more pronounced for larger molecules. Our proposal is applicable to any molecular Hamiltonian in quantum chemistry and is expected to expedite the practical utilization of VQE.
Furthermore, the protocols in JBM-VQE can be utilized in various variational quantum algorithms which optimize the expectation values of many Pauli operators, other than VQE.

The paper is organized as follows.
We present the JBM-VQE algorithm in Sec.~\ref{Sec:Algorithm}.
The required number of measurements to estimate the expectation values of the Pauli operators at certain precision is discussed for JBM-VQE and the conventional VQE in Sec.~\ref{Sec:ShotThresholds}.
The numerical comparison between our proposed JBM-VQE and the conventional VQE on various molecular Hamiltonians is presented in Sec.~\ref{Sec:Results}, demonstrating the acceleration of JBM-VQE.
We discuss several aspects of JBM-VQE in Sec.~\ref{Sec:Discussion}.
Finally, we summarize the paper and provide an outlook in Sec.~\ref{Sec:Conclusions}.

\section{Algorithm}\label{Sec:Algorithm}

In this section, we describe the algorithm of JBM-VQE.
We first explain our target Hamiltonian and the joint Bell measurement.
We then explain the algorithm of JBM-VQE.

\subsection{Setup}
We focus on the quantum chemistry Hamiltonian in the second-quantized form, given by
\begin{equation}
    H_\mathrm{f} = \sum_{i,j=1}^{n} h_{i j} c_i^{\dagger} c_j+\frac{1}{2} \sum_{i,j,k,l=1}^{n} V_{i j k l} c_i^{\dagger} c_j^{\dagger} c_l c_k,
    \label{eq: chem Ham}
\end{equation}
where $c_i \, (c_i^\dag)$ is an annihilation (creation) operator of an electron labeled by $i = 1, 2, \cdots, n$ satisfying the canonical anti-commutation relation $\{c_i, c_j\} = \{c_i^\dag, c_j^\dag\} = 0, \{c_i, c_j^\dag\} = \delta_{ij}$, and $h_{ij} \, (V_{ijkl})$ is the scalar related to the so-called one-electron (two-electron) integrals~\cite{Szabo2012, Helgaker2013}.
This Hamiltonian is mapped to a qubit representation by fermion-qubit mappings such Jordan-Wigner mapping~\cite{Jordan1928}, parity mapping~\cite{Seeley2012}, or Bravyi-Kitaev mapping~\cite{Bravyi2002}, as follows:
\begin{equation}
H = \sum_{j=1}^M \lambda_j P_j,
\label{eq: Ham}
\end{equation}
where $\lambda_j \in \mathbb{R}$ is a coefficient, $P_j$ is an $n$-qubit Pauli operators $P_j \in \{I,X,Y,Z\}^{\otimes n}$, and $M = \order{n^4}$ is the number of the Pauli operators.
Throughout this paper, we omit the identity term $I^{\otimes n}$ in the Hamiltonian and assume $P_j \neq I^{\otimes n}$.
Similar to the conventional VQE, JBM-VQE is based on the ansatz quantum state:
\begin{equation}
 \ket{\psi(\bmth)} = U(\bmth) \ket{0},
\end{equation}
where $U(\bmth)$ is a parameterized quantum circuit with parameters $\bmth = (\theta_1, \cdots, \theta_{N_\theta})$.
Our objective is to minimize the energy expectation value:
\begin{equation}
 E(\bmth) := \ev{H}_{\bmth} = \sum_{j=1}^M \lambda_j \ev{P_j}_{\bmth},
\end{equation}
where $\ev{\cdots}_{\bmth} := \ev{\cdots}{\psi(\bmth)}$, with respect to the parameters $\bmth$.

\subsection{Joint Bell measurement \label{subsec:JBM}}
The joint Bell measurement is an entangling operation in quantum information processing that allows one to distinguish between the four maximally entangled Bell states. It has profound implications for quantum computation, e.g., enabling the determination of the absolute values of expectation values of all $4^n$ Pauli operators for an $n$ qubit state~\cite{Huang2022, huang2021, Jiang2020optimalfermionto}.
It requires a $2n$-qubit system comprising two identical $n$ qubit systems, denoted as $A$ and $B$.
We prepare a $2n$ qubit state,
\begin{equation}
\ket{\Psi(\bmth)} := \ket{\psi(\bmth)}_A \otimes \ket{\psi(\bmth)}_B = (U(\bmth) \otimes U(\bmth)) \ket{0}_A \otimes \ket{0}_B,
\end{equation}
and apply a Controlled-NOT (CNOT) gate with one qubit, $A$, as the control and the other, $B$, as the target. Then we apply a Hadamard gate on the control qubit and eventually measure both qubits in the computational basis (see Fig.~\ref{fig:schematic}).
For each pair of qubits, measuring the state in the computational basis results in the projective measurement onto the Bell basis, 
\begin{equation}
    \begin{aligned}
\ket{\Phi_1^\pm} & =\frac{1}{\sqrt{2}}\left(\ket{0}_A \otimes\ket{0}_B \pm \ket{1}_A \otimes\ket{1}_B\right), \\
\ket{\Phi_2^\pm} & =\frac{1}{\sqrt{2}}\left(\ket{0}_A \otimes\ket{1}_B \pm \ket{1}_A \otimes\ket{0}_B\right).
\end{aligned}
\end{equation}
These basis states are common eigenstates of Pauli operators $X_A \otimes X_B$, $Y_A \otimes Y_B$, $Z_A \otimes Z_B$. Therefore, the measurement for all $2n$ qubits in Fig.~\ref{fig:schematic} constitutes the projective measurement on simultaneous eigenstates of all $2n$ qubit Pauli operators in the form of $P_j \otimes P_j \: (P_j \in \{I,X,Y,Z\}^{\otimes n})$, whose total number is $4^n$.
Consequently, for any $n$ qubit Pauli operators $P_j$, we can estimate 
\begin{align}
 \ev{(P_j \otimes P_j)}{\Psi(\bmth)} = \ev{P_j}_{\bmth}^2
\end{align}
from the measurement outcomes of the single quantum circuit in Fig.~\ref{fig:schematic}.
We denote the estimated value as $\widetilde{\ev{P_j}_{\bmth}^2}$.
The absolute value of $\abs{\ev{P_j}_{\bmth}}$ is then estimated as
\begin{equation}
 \wt{\abs{\ev{P_j}_{\bmth}}} = \sqrt{ \max\{0,  \widetilde{\ev{P_j}_{\bmth}^2} \}}.
 \label{eq: estimate abs}
\end{equation}
We refer to this protocol to estimate the absolute values of expectation values of all $4^n$ Pauli operators as the \textit{joint Bell measurement}.
We observe that the estimate given by Equation \eqref{eq: estimate abs} is biased for a finite number of measurement shots $m$ due to the non-linearity of the square root and the maximum functions although the bias vanishes as $m$ grows.
As shown in Appendix~\ref{appsec:bias}, the bias scales as
\begin{equation} \label{eq:bias scaling}
 \begin{aligned}
 B(\wt{\abs{\ev{P_j}_{\bmth}}}) &= \mathbb{E}^{(m)}(\wt{\abs{\ev{P_j}_{\bmth}}}) -  \abs{\ev{P_j}_{\bmth}} \\
 & \sim 
 \begin{cases}
     m^{-1/4} & (\ev{P_j}_{\bmth} = 0) \\
     -m^{-1}  & (\ev{P_j}_{\bmth} \neq 0)
 \end{cases},
 \end{aligned} 
\end{equation}
where $\mathbb{E}^{(m)}(\cdots)$ signifies the expectation value over $m$ shots measurement.


\subsection{JBM-VQE algorithm}
In the JBM-VQE algorithm, we estimate the energy $E(\bmth)$ and its gradient $\pdv{E(\bmth)}{\bmth}$ by decomposing the expectation value $\ev{P_j}_{\bmth}$ into its sign,
\begin{align}
s_j(\bmth) := \mr{sgn} \left( \ev{P_j}_{\bmth} \right) := \begin{cases}
 +1 & (\mr{if } \ev{P_j}_{\bmth} \geq 0) \\
 -1 & (\mr{if } \ev{P_j}_{\bmth} < 0) 
\end{cases},
\end{align}
and its absolute value $\abs{\ev{P_j}_{\bmth}}$.
We employ the following two subroutines to estimate the energy and gradient in the algorithm.

\textbf{Subroutine 1.}
The first subroutine takes the parameters $\bmth$ as its inputs.
In this subroutine, we first estimate the signs $s_j(\bmth)$ by evaluating the expectation values $\ev{P_j}_{\bmth}$ themselves with the standard measurement strategy using $n$ qubits, as done in the conventional VQE.
One naive way to estimate the signs is to perform the projective measurement of each $P_j$ with $\order{n^4}$ distinct measurement circuits, subsequently estimating the sign $s_j(\bmth)$ via the majority vote of its $\pm 1$ result.
The estimated sign is denoted as $\widetilde{s_j(\bm{\theta})}$, and the total number of shots (repetitions of quantum circuit executions) to estimate all $s_j(\bm{\theta})$ is represented as $m_S^{\text{tot}}$. Following this, the joint Bell measurement using $2n$ qubits is utilized to approximate the absolute values of the expectation values $\abs{\ev{P_j}_{\bmth}}$, resulting in their estimates $\wt{\abs{\ev{P_j}_{\bmth}}}$.
The number of shots for the joint Bell measurement is denoted as $m$.
The estimate of the energy (expectation value of the Hamiltonian) is constructed as
\begin{equation}
    \wt{E(\bmth)} = \sum_j \lambda_j \wt{s_j(\bmth)} \wt{ \abs{\ev{P_j}_{\bmth}} }.
    \label{eq: enegy est. with measured signs}
\end{equation}
Additionally, we estimate the gradient of $E(\bmth)$ by using the so-called parameter shift rule~\cite{ Mitarai2018, Schuld2019}, mathematically expressed in the simplest case as
\begin{equation} \label{eq: parameter shift rule}
\pdv{\ev{P_j}_{\bmth}}{\theta_l}= \frac{1}{2 \sin \alpha} \left( \ev{P_j}_{\bmth^{(l)}_+} - \ev{P_j}_{\bmth^{(l)}_-} \right),
\end{equation}
where ${\bmth}^{(l)}_\pm = \bmth \pm \alpha \bm{\delta}_l$, $\bm{\delta}_l$ is a unit vector with only the $l$-th component non-zero, and $\alpha \in \mathbb{R}$ is a fixed constant. 
We take $\alpha=\pi/4$ in the numerical calculation in Sec.~\ref{Sec:Results}.
Analogous to the energy estimation, we use the standard measurement strategy to estimate the signs $s_j(\bmth^{(l)}_\pm)$ using the quantum states $\ket{\psi(\bmth^{(l)}_\pm)}$, which may require $\order{n^4}$ distinct quantum circuits in a naive way.
The absolute values $\abs{\ev{P_j}_{\bmth^{(l)}_\pm}}$ are then estimated with the joint Bell measurement for the states $\ket{\psi(\bmth^{(l)}_\pm)}$.
The gradient of the energy is estimated by
\begin{align}
 \wt{\pdv{E(\bmth)}{\theta_l}} = \sum_j \frac{\lambda_j}{2\sin \alpha}
 \left( \wt{s_j(\bmth^{(l)}_+)} \wt{\abs{\ev{P_j}_{\bmth^{(l)}_+}}} - \wt{s_j(\bmth^{(l)}_-)} \wt{\abs{\ev{P_j}_{\bmth^{(l)}_-}}}\right).
  \label{eq: grad est. with measured signs}
\end{align}
The estimates of the energy~\eqref{eq: enegy est. with measured signs} and the gradient~\eqref{eq: grad est. with measured signs} are the outputs of this subroutine.

\textbf{Subroutine 2.}
The second subroutine takes the parameters $\bmth$ and a set of guessed signs
\begin{align}
\{t_j\}_{j=1}^M, \: & \{t^{(l=1)}_{j,+}\}_{j=1}^M, \cdots, \{t^{(l=N_\theta)}_{j,+}\}_{j=1}^M, \nonumber \\
& \{t^{(l=1)}_{j,-}\}_{j=1}^M, \cdots, \{t^{(l=N_\theta)}_{j,-}\}_{j=1}^M,
\label{eq: signs for sub2}
\end{align}
as inputs ($t_j, t^{(l)}_{j, \pm} = \pm 1$).
In this subroutine, we estimate only the absolute values ($\abs{\ev{P_j}_{\bmth}}, \abs{\ev{P_j}_{\bmth^{(l)}_+}}$ and $\abs{\ev{P_j}_{\bmth^{(l)}_-}}$) by the joint Bell measurement.
The energy and the gradient are estimated by
\begin{align}
\wt{E(\bmth)} &= \sum_j \lambda_j t_j \wt{ \abs{\ev{P_j}_{\bmth}} }.
\label{eq: enegy est. with given signs}
\\
  \wt{\pdv{E(\bmth)}{\theta_l}} &= \sum_j \frac{\lambda_j}{2\sin \alpha}
 \left( t^{(l)}_{j,+} \wt{\abs{\ev{P_j}_{\bmth^{(l)}_+}}} - t^{(l)}_{j,+} \wt{\abs{\ev{P_j}_{\bmth^{(l)}_-}}}\right).
 \label{eq: grad est. with given signs}
\end{align}
These two estimates are outputs of the second subroutine.

The JBM-VQE algorithm is described in Algorithm~\ref{alg:JBM-VQE}.
Let us assume that the parameters are $\bmth$ in the $n_\mr{iter}$-th iteration ($n_\mr{iter}=0,1,\cdots$) of the algorithm.
When $n_\mr{iter}$ is a multiple of $T_S$, we invoke subroutine 1 to estimate the energy and gradient, $\wt{E(\bmth)}$ [Eq.~\eqref{eq: enegy est. with measured signs}] and $\wt{\pdv{E(\bmth)}{\bmth}}$ [Eq.~\eqref{eq: grad est. with measured signs}], respectively.
Importantly, we also record the estimates of the signs $\wt{s_j(\bmth)}, \wt{s_{j,\pm}^{(l)}(\bmth)}$ for $j=1,\cdots,M$ and $l=1,\cdots,N_\theta$.   
When $n_\mr{iter}$ is not a multiple of $T_S$,  we invoke subroutine 2 with utilizing the pre-recorded signs (obtained at some past iteration) as the guessed signs~[Eq.\eqref{eq: signs for sub2}].
In other words, we estimate the energy and the gradient by Eqs.~\eqref{eq: enegy est. with given signs}\eqref{eq: grad est. with given signs} with performing only the joint Bell measurement that estimates the absolute values of the Pauli expectation values.
Then we update the parameters by the gradient decent method $\bmth' = \bmth - \eta \wt{\pdv{E(\bmth)}{\bmth}}$, where $\eta$ is a learning rate. 
It is worth noting that the gradient descent is not the only choice in the JBM-VQE and other sophisticated optimization algorithms can be employed (see the discussion in Sec.~\ref{Sec:Discussion}).

Several remarks regarding our JBM-VQE algorithm are in order.
Firstly, this algorithm relies on the expectation that the signs of the Pauli expectation values ($\ev{P_j}_{\bmth}$ and $\ev{P_j}_{\bmth^{(l)}_\pm}$) do not change frequently during the optimization process.
Subroutine 2 consists of the joint Bell measurement that uses only $(2N_\theta + 1)$ quantum circuits and may typically require fewer measurement shots to estimate the absolute values of the Pauli expectation values than the conventional VQE.
For this reason, we expect a reduction in the total number of shots in JBM-VQE compared with the conventional VQE.
Secondly, the joint Bell measurement has a bias on its estimates, causing the energy and gradient estimated in both subroutines 1 and 2 to exhibit bias. Nevertheless, this bias vanishes with the increase of the number of measurement shots and remains a minor concern since it is relatively small compared to the required energy precision, especially during the early optimization stages. (This topic is further elaborated upon in Sec.~\ref{Sec:Discussion} and Appendix~\ref{appsec:bias}). The rough criteria of the number of shots for realizing a certain precision of the estimated expectation values are discussed in Sec.~\ref{Sec:ShotThresholds}. 
Thirdly, if we use a $2n$-qubit system just as two independent copies of the original $n$-qubit system and conduct the conventional VQE,  $m$ executions of the circuits are equivalent to $2m$ shots in the original system. Consequently, our JBM-VQE algorithm must surpass the conventional VQE by at least a factor of two concerning the number of shots, and it is actually realized in the numerical simulation in Sec.~\ref{Sec:Results}. Lastly, the sign-updating period $T_S$ influences the efficiency of JBM-VQE. A larger period leads to fewer shots required for optimization, albeit with the trade-off of less accurate energy and gradient estimates. While there is no a priori criterion for determining $T_S$, it can be set manually or adaptively by monitoring the optimization history.


\begin{algorithm*}[t]
\caption{JBM-VQE. \label{alg:JBM-VQE}}
\SetKwInOut{Return}{Return}
\KwIn{Hamiltonian $H=\sum_j\lambda_jP_j$, variational circuit $U(\boldsymbol{\theta})$, number of shots for the joint Bell measurement $m$, sign-updating period $T_S$, number of shots for sign-updating $m_S^\mr{tot}$, and learning rate $\eta$.}
\KwOut{Parameters $\bmth_{\mathrm{opt}}$ which approximate the ground state of $H$ by $\ket{\psi(\bmth_{\mathrm{opt}})}=U(\bmth_{\mathrm{opt}})\ket{0}$ and the estimated optimal energy $\wt{E(\bmth)}$.}
Initialize $\bmth$ and set $n_\mathrm{iter}=0$\;
\While{energy estimate $\wt{E(\bmth)}$ has not converged}{
  \If{$n_\mathrm{iter} \, \mathrm{mod} \, T_S = 0$}{
        Call subroutine 1 with input $\bmth$, and obtain the estimates of the energy~[Eq.~\eqref{eq: enegy est. with measured signs}] and the gradient [Eq.~\eqref{eq: grad est. with measured signs}]\;
        Record the signs $\wt{s_j(\bmth)}, \wt{s_{j,\pm}^{(l)}(\bmth)}$ as $t_j, t^{(l)}_{j,\pm}$\; 
      }
  \If{$n_\mathrm{iter} \, \mathrm{mod} \, T_S \neq 0$}{
        Call subroutine 2 with input $\bmth$ and the guessed signs $t_j, t^{(l)}_{j,\pm}$ at some past iteration, and obtain the estimates of the energy~[Eq.~\eqref{eq: enegy est. with given signs}] and the gradient [Eq.~\eqref{eq: grad est. with given signs}]\;
      }
  $\bmth \leftarrow \bmth - \eta \wt{\pdv{E(\bmth)}{\bmth}}$\;
  $n_\mathrm{iter} \leftarrow n_\mathrm{iter} + 1$\;
  }
 Return $\bmth$ and  $\wt{E(\bmth)}$\;
\end{algorithm*}

\section{Shot Thresholds}\label{Sec:ShotThresholds}

In both JBM-VQE and the conventional VQE, the energy estimate exhibits finite statistical fluctuation due to the limited number of measurement shots. This occurs even without any noise present in quantum devices. To facilitate a fair comparison between JBM-VQE and the conventional VQE, it is essential to establish a common criterion ensuring that both methods display the same level of fluctuation.
In this section, we discuss such a criterion by investigating the number of shots required to estimate an expectation value of a single Pauli operator with a fixed level of accuracy.
We formulate the number of shots to estimate the expectation value with certain accuracy and probability, and numerically calculate the actual numbers.
The number established here is utilized in Sec.~\ref{Sec:Results}, where numerical demonstrations of JMB-VQE are performed for quantum chemistry Hamiltonians of small molecules.

\subsection{Shot threshold for the conventional VQE}
Let us define the number of shots to estimate the expectation value of a single Pauli operator with the projective measurement, which is the standard measurement strategy for the conventional VQE.
For a given $n$-qubit state $\ket{\psi}$ and an $n$-qubit Pauli operator $P$, the probability of estimating $\ev{P} := \ev{P}{\psi}$ within an additive error $\tau_{\mr{th}}$ under the $m$-shot projective measurement of $P$ is given by
\begin{align}\label{probability_VQE}
 p(m, \tau_{\mr{th}}, \ev{P}) &= \sum_{x=x_\mr{min}}^{x_\mr{max}} 
\binom{m}{x} \left(\frac{1+\ev{P}}{2}\right)^x \left(\frac{1-\ev{P}}{2}\right)^{m-x}, \\
x_\mr{min} &= \mr{max} \Big\{ 0, \left\lceil \frac{ m(1+\ev{P}-\tau_{\mr{th}}) }{2} \right\rceil \Big\}, \nonumber \\ 
x_\mr{max} &= \mr{min} \Big\{ m, \left\lfloor \frac{ m(1+\ev{P}+\tau_{\mr{th}}) }{2} \right\rfloor \Big\}, \nonumber
\end{align}
where $\lceil ... \rceil$ ($\lfloor ... \rfloor$) is the ceiling (floor) function of integers. Here the expectation value $\ev{P}$ itself only matters to calculate the threshold and there is no dependence on the number of qubits. This is partly because we consider the single Pauli operator to define the threshold.
Since there are various Pauli operators included in the Hamiltonian, we consider the averaged probability for estimating the expectation value within an additive error $\tau_{\mr{th}}$,
\begin{align}
 p^{(\mr{av})}(m, \tau_{\mr{th}}) = \frac{1}{2} \int_{-1}^1 dy \, p(m, \tau_{\mr{th}}, y).
\end{align}
We then define the standard measurement (SM) shot threshold as follows:
\begin{definition}[SM shot threshold] \label{Def:Shot_threshold}
The standard measurement (SM) shot threshold $m^{\mr{SM}}_{\mr{th}}(\tau_{\mr{th}}, p_{\mr{th}})$ is defined as 
\begin{equation}
m_{\mr{th}}^{\mr{SM}}(\tau_{\mr{th}}, p_{\mr{th}}) := \min\{m\in\mathbb{Z}^+ \mid p^{(\mr{av})}(m, \tau_{\mr{th}}) \geq p_{\mr{th}}\}
\end{equation}
\end{definition}
The SM shot threshold $m_{\mr{th}}^{\mr{SM}}(\tau_{\mr{th}}, p_{\mr{th}})$ indicates the minimum number of the shots of the projective measurement of $P$ to estimate $\ev{P}$ within an additive error $\tau_{\mr{th}}$ with a probability at least $p_{\mr{th}}$, where the expectation value $\ev{P}$ is averaged in the uniform distribution for $[-1, 1]$.
We leverage $m_{\mr{th}}^{\mr{SM}}(\tau_{\mr{th}}, p_{\mr{th}})$ to determine the number of shots in numerical simulation of the conventional VQE in Sec.~\ref{Sec:Results}.
We note that the value of $\ev{P}$ may cluster around $0$ if we consider random states in the Hilbert space, e.g., Haar random states, but we employ the uniform distribution because the ground states of quantum chemistry Hamiltonians are not random states and various values of $\ev{P}$ may appear. Thus, while the threshold would differ for other distributions, we have chosen the uniform distribution for its relevance and simplicity as a baseline for this study. See Appendix~\ref{Appendix: Ground-state distribution} for discussions on another distribution of $\ev{P}$.

\begin{figure}[tbh]
	\includegraphics[width=0.35\textwidth]{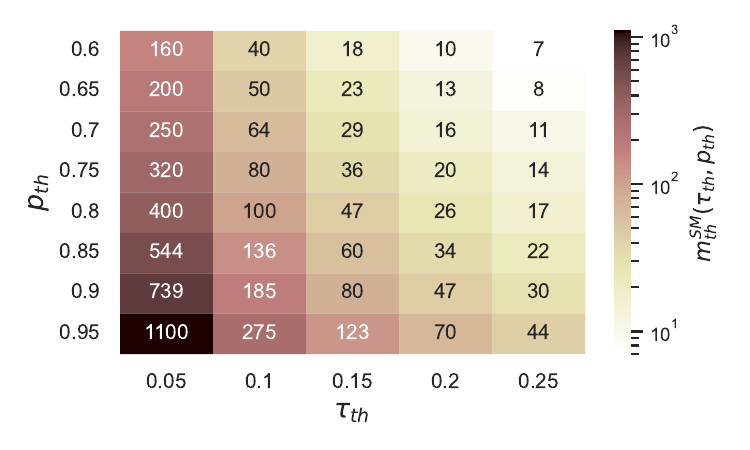}
	\caption{Values of the standard measurement (SM) shot threshold $m^{\mr{SM}}_{\mr{th}}(\tau_{\mr{th}}, p_{\mr{th}})$ for different estimation errors $\tau_{\mr{th}}$ and probabilities $p_{\mr{th}}$.}
    \label{tab:SM_threshold}
\end{figure}

Finally, we evaluate actual numerical values of $m_{\mr{th}}^{\mr{SM}}(\tau_{\mr{th}}, p_{\mr{th}})$ for various $\tau_{\mr{th}}$ and $p_{\mr{th}}$.
For a given $m$, we calculate $p^{\mr{ave}}(m, \tau_{\mr{th}})$ by approximating the integral through numerical integration, taking 2000 uniformly-spaced points of $\ev{P}$ in the interval $[-1,1]$.
The results are presented in Fig.~\ref{tab:SM_threshold}.


\subsection{Shot threshold for JBM-VQE}
In JBM-VQE, the estimation of the absolute value $\ev{P}$ and the sign $s := \mr{sign}(\ev{P})$ for a given state $\ket{\psi}$ and a Pauli operator $P$ is performed differently.
The absolute value is estimated using joint Bell measurements, while the sign is typically determined through the projective measurement of $P$.
We consider the number of shots to estimate each of them within certain error and probability. 

\begin{figure}[tbh]
	\includegraphics[width=0.35\textwidth]{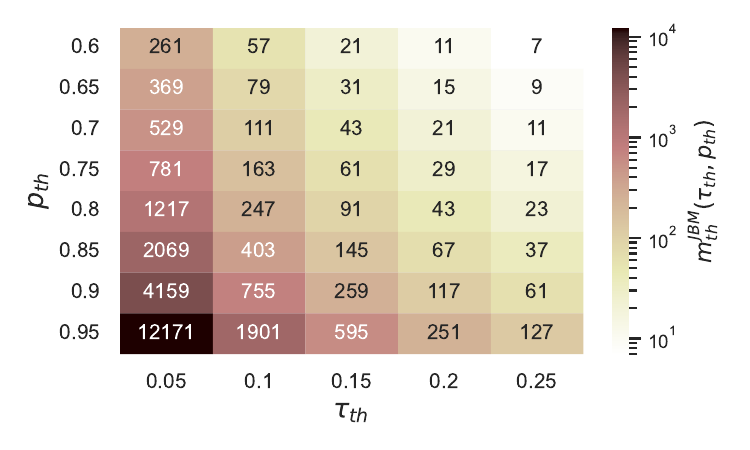}
	\caption{Values of the joint Bell measurement (JBM) shot threshold $m^{\mr{JBM}}_{\mr{th}}(\tau_{\mr{th}}, p_{\mr{th}})$ for different estimation errors $\tau_{\mr{th}}$ and probabilities $p_{\mr{th}}$.}
    \label{tab:JBM_threshold}
\end{figure}

First, let us define the number of shots necessary for accurately estimating the absolute value $\ev{P}$.
Since the joint Bell measurement provides the expectation values of $P \otimes P$, the probability of determining $|\langle P \rangle|$ within an error $\tau_{\mr{th}}$ is calculated as follows:
\begin{align}\label{probability_JBMVQE}
  & q(m, \tau_{\mr{th}}, \ev{P}) \nonumber \\
 =& \sum_{x \in \mathbb{X}}
\binom{m}{x} \left(\frac{1+\ev{P}^2}{2}\right)^x \left(\frac{1-\ev{P}^2}{2}\right)^{m-x},
\end{align}
where $\mathbb{X}$ is a set of integers in $0 \leq x \leq m$ that satisfies
\begin{align}
    \abs{ \sqrt{\max \big\{0, \frac{2x}{m}-1\big\}} - \abs{\ev{P}}} \leq \tau_{\mr{th}}
\end{align}
(see Eq.~\eqref{eq: estimate abs}).
Similar to SM shot threshold, we average the probability $q(m, \tau_{\mr{th}}, \ev{P})$ as 
\begin{align}
 q^{(\mr{av})}(m, \tau_{\mr{th}}) = \frac{1}{2} \int_{-1}^1 dy \, q(m, \tau_{\mr{th}}, y).
\end{align}
We can now define JBM shot threshold:
\begin{definition}[JBM Shot Threshold]\label{Def:JBM_Shot_threshold}
JBM shot threshold $m^{\mr{JBM}}_{\mr{th}}(\tau_{\mr{th}}, p_{\mr{th}})$ is defined as 
\begin{equation}
m_{\mr{th}}^{\mr{JBM}}(\tau_{\mr{th}}, p_{\mr{th}}) := \min\{m\in\mathbb{Z}^+ \mid p^{(\mr{av})}(m, \tau_{\mr{th}}) \geq p_{\mr{th}}\}
\end{equation}
\end{definition}
This means that the JBM shot threshold $m^{\mr{JBM}}_{\mr{th}}(\tau_{\mr{th}}, p_{\mr{th}})$ indicates the minimal number of shots needed to estimate the absolute value of the expectation value $\ev{P}$ within an additive error $\tau_{\mr{th}}$ with probability $p_{\mr{th}}$.
Numerical values of $m^{\mr{JBM}}_{\mr{th}}(\tau_{\mr{th}}, p_{\mr{th}})$ are again calculated by 2000 points of $\ev{P}$ that are uniformly-spaced in $[-1,1]$ and summarized in Fig.~\ref{tab:JBM_threshold}.


\begin{figure}[tbh]
	\includegraphics[width=0.48\textwidth]{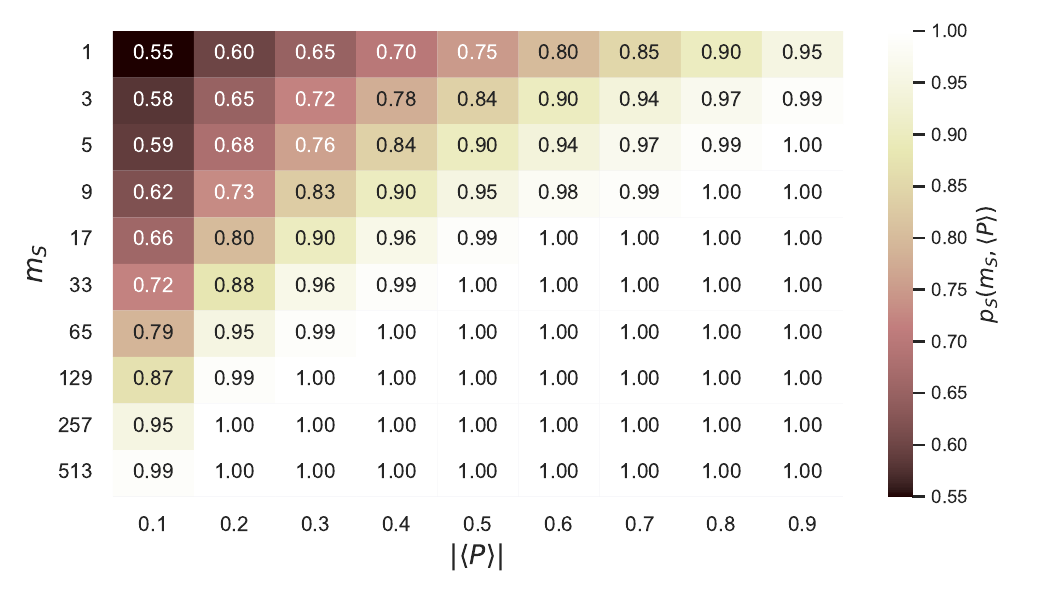}
	\caption{Values of the probability $p_S(m_S, \ev{P})$ for different measurement shots $m_S$ and absolute values of Pauli expectation values $|\langle P \rangle|$.}
    \label{tab:sign_threshold}
\end{figure}

Next, we examine the estimation of the sign $s$ in JBM-VQE that is performed through a majority vote of the results of the projective measurement of $P$.
When the number of shots is an odd integer $m_S$, the probability of estimating $s$ correctly is
\begin{align}
 p_S(m_S, \ev{P}) &= \sum_{x \in \mathbb{X}_S} 
\binom{m_S}{x} \left(\frac{1+\ev{P}}{2}\right)^x \left(\frac{1-\ev{P}}{2}\right)^{m_S-x}, \\
\mathbb{X}_S & =
\begin{cases}
\{ \frac{m_S+1}{2}, \frac{m_S+3}{2}, \cdots, m_S \} & (\ev{P} \geq 0) \\
\{ 0, 1, \cdots, \frac{m_S-1}{2} \} & (\ev{P} < 0)
\end{cases}
\end{align}
The numerical values of $p_S(m_S, \langle P \rangle)$ are presented in Fig.~\ref{tab:sign_threshold}.
We observe that even for a relatively small number of shots such as $m_S = 17$, the probability $p'$ is as high as $\geq 0.8$ for $\ev{P} \ge 0.2$.
In our numerical simulations presented in the next section, we take these values into account for determining the value of $m_S$ in the JBM-VQE algorithm.

Finally, it is worthwhile to point out that $m_\mr{th}^\mr{SM}$ and $m_S$ considered here are defined for a single Pauli operator $P$, so if there are $M$ Pauli operators, it would require $M \cdot m_\mr{th}^\mr{SM}$ and $M \cdot m_S$ shots to estimate all expectation values and their sign.
Similarly, it is shown~\cite{huang2021} that $\order{\log(M)/\epsilon^4}$ measurements are need to estimate $\abs{\ev{P_1}}, \cdots, \abs{\ev{P_M}}$ within the error $\epsilon$ simultaneously,  so it would require $\log(M) \cdot m_\mr{th}^\mr{JBM}$ to estimate all absolute values of them.


\section{Numerical demonstration}\label{Sec:Results}
\begin{figure}[t]
	\includegraphics[width=0.49\textwidth]{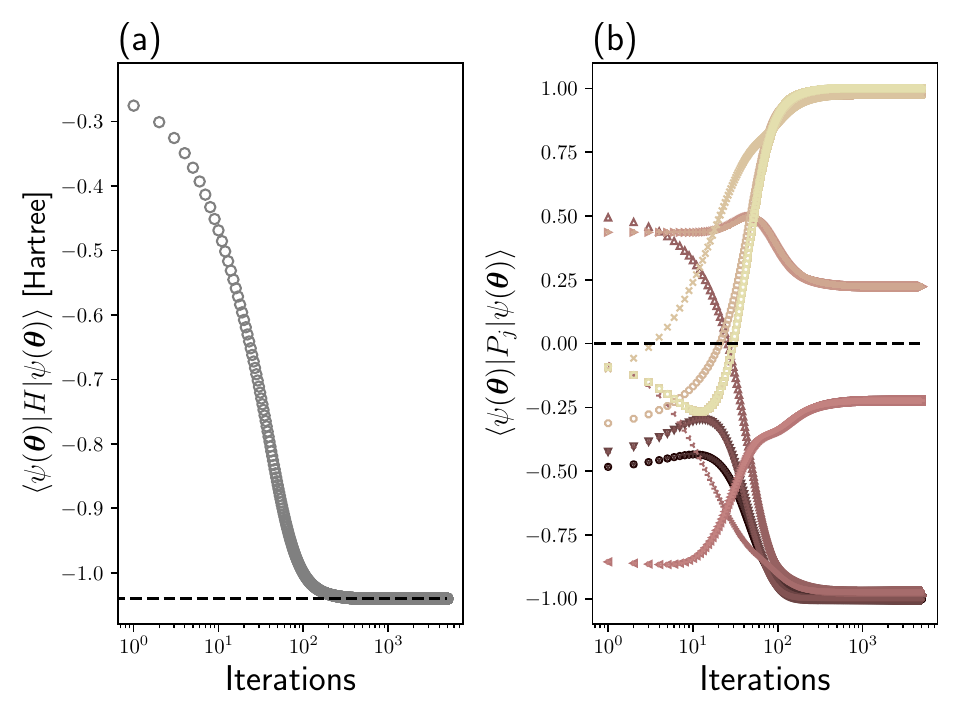}
	\caption{Expectation values of (a) the Hamiltonian $H$ and (b) the Pauli operators $P_j$ included in the Hamiltonian during the VQE optimization for finding the ground state of the \ce{H2} molecular Hamiltonian. There are 14 non-identity Pauli operators included in the Hamiltonian $H$.}
    \label{fig:Exps}
\end{figure}

In this section, we present numerical demonstrations of JBM-VQE for finding the ground states of quantum chemistry Hamiltonians.
We first examine an example where the signs of Pauli expectation values, $s_j(\bmth)$, do not undergo frequent flipping during the course of standard VQE optimization.
This observation leads to the anticipation that JBM-VQE can reduce the number of shots to optimize the parameters with a specified level of accuracy.
We then compare JBM-VQE and the conventional VQE by taking various small molecules as examples.
The parameter optimization in JBM-VQE proceeds more rapidly than in the conventional VQE, according to the metric we have established for the early stage of the optimization.
The advantage of JBM-VQE becomes increasingly apparent as system sizes (the number of qubits) grow larger.

\subsection{Signs of Pauli expectation values during parameter optimization}
We conduct a numerical simulation of the conventional VQE to prepare the ground state of the $\ce{H2}$ molecular Hamiltonian whose bond distance is 0.74\AA.
The quantum chemistry Hamiltonian of the form~\eqref{eq: chem Ham} is constructed by using the Hartree-Fock molecular orbitals with the STO-3G basis set.
The Hamiltonian is then mapped to the qubit form~\eqref{eq: Ham} via the Jordan-Wigner transformation.
The resulting qubit Hamiltonian consists of $n=4$ qubits and 14 non-identity Pauli operators.
The construction of the Hamiltonian was implemented using the numerical libraries PySCF~\cite{Sun2018} and OpenFermion~\cite{McClean2020}.
For the VQE ansatz state, we employ the symmetry-preserving ansatz~\cite{gard2020efficient,ibe2022calculating} $\ket{\psi(\bmth)} = U(\bmth)\ket{n_e}$, where $U(\bmth)$ represents a variational quantum circuit depicted in Fig.~\ref{fig:Circuit} of Appendix~\ref{appsec:numerics} and $\ket{n_e}$ is a computational basis state $\ket{0\cdots 0 1\cdots 1}$ with $n-n_e$ ``0"s and $n_e$ ``1"s with $n_e$ denoting the number of electrons (for \ce{H2} molecule, $n_e=2$).
It has eight parameters in total, with initial values sampled randomly from $[0,\pi/5]$. Here we choose to start from a random initial state instead of the Hartree-Fock state. This decision is grounded in the understanding that the Hartree-Fock initialization might limit the VQE's ability to explore more correlated states and lead VQE to converge to a local minimum.
Furthermore, it is recognized that for certain molecules such as diatomic molecules (e.g., the nitrogen \ce{N2}) with large bond distances, the HF state is not a good approximation of the true ground state. The circuit parameters are updated using the gradient descent method with a learning rate $\eta = 0.02$.
We simulate the energy expectation values $E(\bmth)=\ev{H}{\psi(\bmth)}$ exactly without assuming any statistical error and noise sources by the numerical library Qulacs~\cite{Suzuki2021}.

The result is presented in Fig.~\ref{fig:Exps}.
After 3000 iterations, the VQE algorithm successfully finds the exact ground state.
Importantly, we observe that the signs of the expectation values of the most Pauli operators $\ev{P_j}{\psi(\bmth)}$ either remain unchanged or exhibit only a single change during the VQE optimization.
This observation motivates us to accelerate VQE by estimating only the absolute values of $\ev{P_j}{\psi(\bmth)}$ in the majority of iterations.
We note that expectation values of some Pauli operators (e.g., $\langle X_1\otimes Y_2\otimes Y_3\otimes X_4\rangle$ and $\langle Y_1\otimes X_2\otimes X_3\otimes Y_4\rangle$) are the same throughout the optimization process due to the ansatz symmetry.


\subsection{Comparison of JBM-VQE with the conventional VQE for various small molecules}
Here we present a numerical comparison of the measurement cost between JBM-VQE and the conventional VQE.
We consider eight molecular systems: \ce{H2}, \ce{H3+}, \ce{H4}, \ce{H5+}, \ce{LiH}, \ce{H2O}, \ce{NH3}, and \ce{BeH2}.
The geometries of these molecules are provided in Table~\ref{tab: geometries} of Appendix~\ref{appsec:numerics}.
For the latter four molecules (\ce{LiH}, \ce{H2O}, \ce{NH3}, and \ce{BeH2}), the active space approximation of four orbitals are taken so that the Hamiltonians become 8-qubit ones.

Similar to the calculations in the previous subsection, we construct quantum chemistry Hamiltonians of the form~\eqref{eq: chem Ham} using Hartree-Fock molecular orbitals with the STO-3G basis set for the eight molecules under consideration.
The Jordan-Wigner transformation is employed to obtain the qubit Hamiltonian.
The symmetry-preserving ansatz (described in Fig.~\ref{fig:Circuit} of Appendix~\ref{appsec:numerics}) is employed for both JBM-VQE and the conventional VQE. The ansatz depths for molecules \ce{H2}, \ce{H3+}, \ce{H4}, \ce{H5+}, \ce{LiH}, \ce{H2O}, \ce{NH3}, and \ce{BeH2} are 2, 3, 8, 14, 3, 5, 6, and 5, respectively. Initial parameters are uniformly sampled from $[0,\pi/5]$ for all cases. 

We evaluate the expectation value $E(\bmth) = \ev{H}{\psi(\bmth)}$ and its gradient as follows.
In JBM-VQE, the signs $\{ s_j(\bmth) \}_{j=1}^M$ and $\{ s_{j}(\bmth^{(l)}_\pm) \}_{j=1}^M$ are estimated by the projective measurement of the Pauli operators $P_j$ included in the Hamiltonian. 
We employ the qubit-wise commuting (QWC) grouping~\cite{Kandala2017, Verteletskyi2020} to make groups of simultaneously-measurable Pauli operators, which requires additional $\order{n}$ one-qubit gates and reduces the number of the total shots for estimation (see Appendix~\ref{appsec:numerics} for details).
We allocate the same number of shots for all generated groups, denoting the number of shots for each group as $m_S$. The absolute values $\{ \abs{\ev{P_j}_{\bmth}} \}_{j=1}^M$ are estimated using the joint Bell measurement with $2n$ qubits, as described in Sec.~\ref{Sec:Algorithm}. The number of shots for the joint Bell measurement is denoted by $m$.
In the conventional VQE, we estimate the energy expectation value by directly estimating $\ev{P_j}{\psi(\bmth)}$ with the projective measurement of $P_j$.
We also employ QWC grouping to reduce the number of shots and allocate the same amount of shots for all groups.
The number of shots for each group is denoted by $m_\mr{VQE}$.
For both JBM-VQE and the conventional VQE, all outcomes of the measurement shots (bitstrings) are simulated without considering any noise sources using the numerical library Qulacs~\cite{Suzuki2021}. 

The learning parameter for the gradient descent is set to $\eta=0.02$ for both JBM-VQE and the conventional VQE.
As mentioned in Sec.~\ref{Sec:Algorithm}, we set $\alpha=\pi/4$ in the parameter shift rule~\eqref{eq: parameter shift rule} to estimate the gradient because of the reason described in Appendix~\ref{appsec:numerics}.
We consider the threshold of $p_{\mr{th}} = 0.9$ and $\tau_{\mr{th}} = 0.05$ for estimating the Pauli expectation values.
According to Figs.~\ref{tab:SM_threshold}, \ref{tab:JBM_threshold} and \ref{tab:sign_threshold}, the numbers of the shots are taken as $m=4159$, $m_S=513$, and $m_\mr{VQE}=739$.
Note that the standard deviations (or fluctuations) in estimating the energy expectation values are not strictly the same between JBM-VQE and the conventional VQE because the values of the thresholds discussed in Sec.~\ref{Sec:ShotThresholds} are for a single Pauli operator.
The sign-updating period of JBM-VQE is fixed at $T_S=30$ for all simulations. For every 200 iterations, the mean estimated energy is computed. Should the energy fail to decrease by a minimum of $0.001$ following $200$ iterations, the optimization process is terminated.

\begin{figure}[t]
	\centering
	\includegraphics[width=0.49\textwidth]{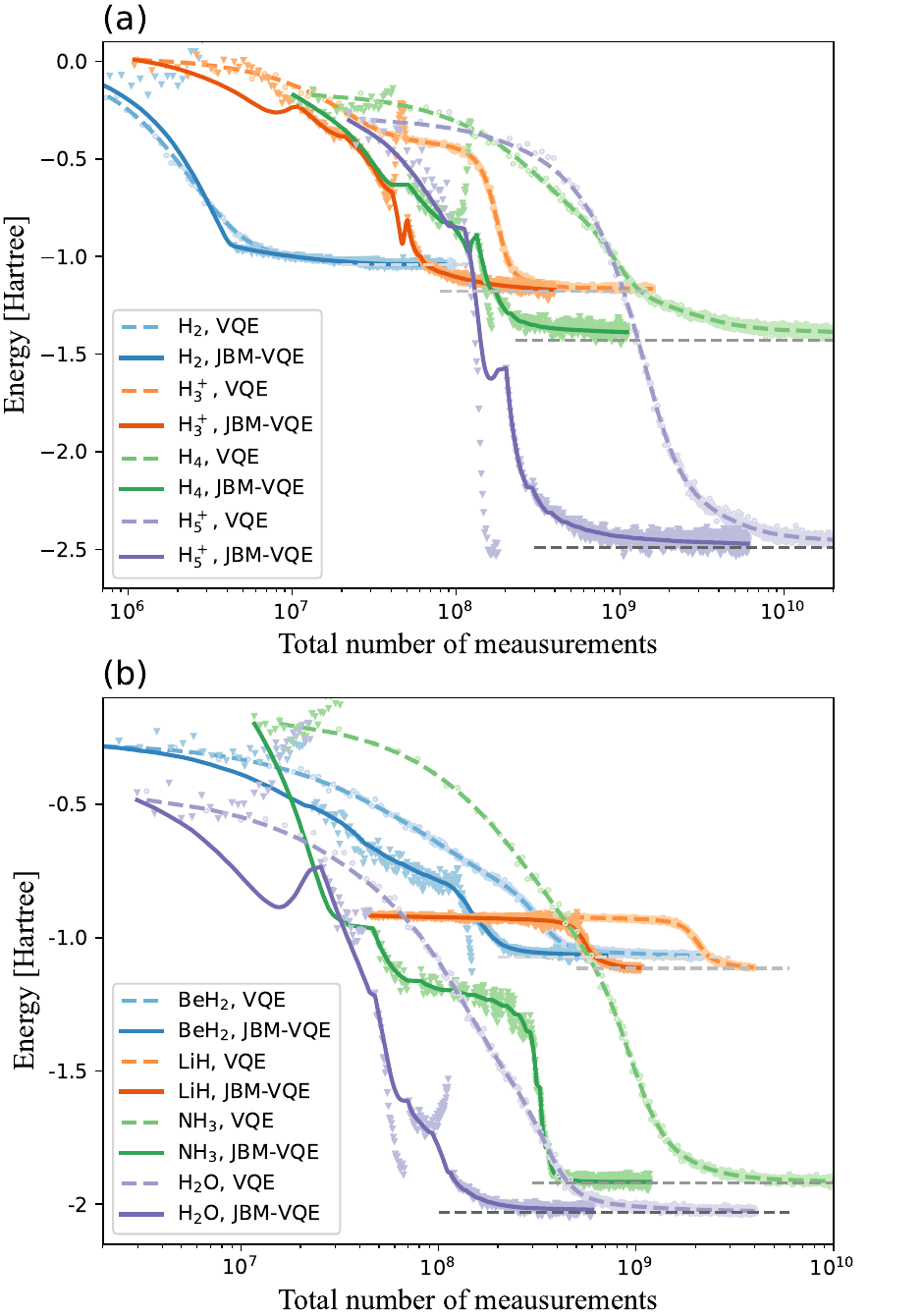}
	\caption{Comparison of optimization curves for the conventional VQE and the JBM-VQE algorithms applied to molecular Hamiltonians of (a) $\ce{H2}$, $\ce{H3+}$, $\ce{H4}$, $\ce{H5+}$ and (b) $\ce{LiH}$, $\ce{H2O}$, $\ce{NH3}$, and $\ce{BeH2}$. The \red{dashed} lines and circles denote the actual energy (exact energy expectation value for the parameters at each iteration) and the estimated energy in the conventional VQE, respectively. The \red{solid} lines and triangles represent the actual energy and the estimated energy in JBM-VQE, respectively. 
 }
	\label{fig:Hn_mol}
\end{figure}

The results are shown in Fig.~\ref{fig:Hn_mol}.
For all molecules, JBM-VQE exhibits a faster decrease of energy with fewer measurement shots.
The sudden jump of the estimated energy in JBM-VQE appears to be due to the update of the signs. The observed bias in our numerical experiment is not so significant to initiate the optimization.
It is also observed that the advantage of JBM-VQE becomes more evident as the system size increases from 4 qubits (\ce{H2}) to 10 qubits (\ce{H5+}).
To quantify the improvement of the  measurement cost of JBM-VQE over the conventional VQE, we count the number of the shots required to achieve a result with energy lower than the Hartree-Fock energy. 
The ratio between such shots for the conventional VQE over JBM-VQE is calculated by averaging the results of at least 50 independent numerical simulations with different initial parameters, which yields ($n$ denotes the number of qubits)
\begin{equation}
    \begin{aligned}
        &\ce{H2}(n=4):1.44, && \ce{H3+}(n=6):3.45,\\
        &\ce{H4}(n=8):9.00, &&\ce{H5+}(n=10):9.24,\\
        &\ce{BeH2}(n=8):3.70, &&\ce{LiH}(n=8):4.60,\\
        &\ce{NH3}(n=8):11.60, &&\ce{H2O}(n=8):2.24.
    \end{aligned}
\end{equation}
These numbers illustrate the reduction of the measurement cost of JBM-VQE.

Finally, we comment on the choice of the Hartree-Fock energy to define the number of shots for optimizing the parameters of the ansatz although the Hartree-Fock energy can be obtained by the initial state of the ansatz without applying $U(\bmth)$.  
This is because the purpose of this numerical illustration is to simply show the reduction of the number of shots in the optimization process of (JBM-)VQE.
Furthermore, it is important to acknowledge that the ansatz selected in this study does not guarantee the accurate representation of the ground-state for the given Hamiltonian. This limitation arises due to the expressibility constraints of the ansatz and make it difficult to define the number of shots for the optimization by using the exact ground state energy.

%

\section{Discussion}\label{Sec:Discussion}
In this section, we discuss several aspects of the JBM-VQE method.

First, JBM-VQE can be expected to present a significant advantage over conventional VQE as the number of qubits $n$ increases, as demonstrated in the hydrogen chain examples from the previous section.
This is because the number of distinct quantum circuits to evaluate the energy expectation value scales at least $\order{n^2}$ (naively $\order{n^4}$) in the conventional VQE while that of JBM-VQE does $\order{1}$ when we skip the evaluation of the sign of the Pauli expectation values.
Although a small number of distinct quantum circuits does not directly imply the efficiency of the evaluation, we anticipate a reduction of the total number of shots in JBM-VQE, as illustrated in the previous section.

Second, we propose using JBM-VQE as an initial optimizer when the accuracy of the energy estimate is not as high, as demonstrated in the numerical simulation in Sec.~\ref{Sec:Results}, because of the following two reasons.
The first reason is that the estimate of the energy in JBM-VQE is biased, i.e., the average of $\wt{E(\bmth)}$ (Eqs.~\eqref{eq: enegy est. with measured signs}\eqref{eq: enegy est. with given signs}) is not equal to the true value $E(\bmth)$ (note that the estimate $\wt{E(\bmth)}$ is a random variable whose probability distribution is determined by that of the outcomes of the joint Bell measurement).
The bias has already been seen in the numerical simulation~(Fig.~\ref{fig:Hn_mol}), where the distribution of the dots (the energy estimates) is not centered at the corresponding line (the exact value of the energy at those parameters) for several molecules.
The second reason is the scaling of the number of shots in the joint Bell measurement with respect to the estimation error.
As shown in Ref.~\cite{huang2021}, the number of shots required to estimate all $M$ absolute values $\abs{\ev{P_1}}, \cdots, \abs{\ev{P_M}}$ with the error $\epsilon$ by the joint Bell measurement is $\order{\log(M)/\epsilon^4}$.
This $\epsilon^{-4}$ scaling can be problematic when we take small $\epsilon$.
These two properties of JBM-VQE can pose challenges when optimizing the ansatz parameters with high accuracy required quantum chemistry, such as the so-called chemical accuracy $10^{-3}$ Hartree.

Thirdly, the JBM-VQE method is specifically designed for NISQ devices where noise is an inevitable factor. The JBM approach inherently generates entanglement between two state copies, necessitating twice the number of gates compared to conventional methods. Considering an error rate per gate denoted as $p$ and a total gate count of $K$, the application of the probabilistic error cancellation~\cite{Temme2017}, a quantum error mitigation strategy, incurs a sampling cost expressed as $e^{4Kp}$~\cite{Cai2022}. In the context of the JBM-VQE, where the gate number $K$ is nearly doubled, this cost scales to $e^{8Kp}$. Consequently, the ratio of these costs is $e^{4Kp}$. Practically, for the usual VQE with the probabilistic error mitigation to work, the value of $Kp$ is limited to at most $\mathcal{O}(1)$, implying that the escalated sampling cost of probabilistic error cancellation in JBM-VQE compared to conventional VQE methods, quantified as $e^{4Kp}$, does not scales with the qubit count $n$.

Lastly, we point out that the JBM-VQE protocol is flexible and can be combined with various variational algorithms which require the evaluation of many Pauli expectation values to optimize the parameters.
For example, sophisticated optimizers like Adam~\cite{kingma2017} or the ones more tailored for VQE~\cite{Kubler2020, arrasmith2020} can be used in JBM-VQE although we employ the plain-vanilla gradient decent in the numerical simulations.
One can simply skip the evaluation of the signs of Pauli expectation values at some iterations (parameter updates) and perform the joint Bell measurement to estimate the energy and its gradient that are fed into the optimizers.

\section{Summary and Outlook}\label{Sec:Conclusions}

In this study, we introduce a protocol designed to accelerate the Variational Quantum Eigensolver (VQE) algorithm for determining the ground state of molecular Hamiltonians. Our approach employs the joint Bell measurement to estimate the energy expectation value and its gradient of $n$ qubit systems with $\order{n^0}=\order{1}$ distinct quantum circuits in the majority of iterations, under the assumption that the signs of Pauli expectation values do not frequently change during optimization. In contrast, the conventional VQE necessitates at least $\mathcal{O}(n^2)$ distinct quantum circuits for estimating the energy and gradient in each iteration. 
We conducted numerical simulations of various small molecular Hamiltonians, demonstrating that our proposed protocol effectively reduces the number of measurement shots required to optimize the ansatz parameters to a certain level. JBM-VQE holds promise for application in a broad range of near-term quantum algorithms that depend on Pauli expectation value estimations.

In future work, it is interesting to apply our protocol to various NISQ algorithms other than the simple VQE presented in this study.
For example, it is possible to combine JBM-VQE with the variants of VQE for excited states~\cite{Nakanishi2019, Parrish2019}, quantum imaginary-time evolution~\cite{Motta2020, McArdle2019, Sun2021, cao2022B}, and algorithmic error mitigation schemes~\cite{Cao2023, Suchsland2021}. There is reason to believe that in protocols involving estimating quantum computed moments $\langle \psi |H^k| \psi \rangle$ like the Lanczos-inspired error mitigation scheme~\cite{Suchsland2021}, variance-VQE (where $k=2$)~\cite{zhang2020}, and variance extrapolation~\cite{Cao2023}, the acceleration from joint Bell measurement become more significant since the number of Pauli terms to be evaluated is of order $\mathcal{O}(n^{4k})$. Furthermore, it would be beneficial to explore the integration of our protocol with two notable strategies - the $\alpha$-VQE~\cite{Wang2019Accelerated} and the parallelized VQE~\cite{Mineh2023Accelerating}. Both of these approaches have demonstrated promising results in improving the efficiency of VQE optimization.

\begin{acknowledgements}
We thank Ridwan Sakidja, Tiglet Besara, Daniel Miller, and Aonan Zhang for helpful discussions.
\end{acknowledgements}

\appendix
\section{Bias in JBM-VQE} \label{appsec:bias}
In this section, we discuss the bias in estimating $\ev{P_j}_{\bmth} = \ev{P_j}{\psi(\bmth)}$ by the joint Bell measurement for a given state $\ket{\psi(\bmth)}$ and Pauli operator $P_j$.
The joint Bell measurement yields measurement outcomes that obey the binomial distribution whose mean is $\ev{P_j}_{\bmth}^2$.
When the number of shots $m$ for the joint Bell measurement is large, the bias of $\wt{\ev{P_j}_{\bmth}}$ can be assessed by approximating the binomial distribution by the normal distribution:
\begin{equation}
\begin{aligned}
    &B(\wt{\abs{\ev{P_j}_{\bmth}}}) = \mathbb{E}^{(m)}(\wt{\abs{\ev{P_j}_{\bmth}}}) -  \abs{\ev{P_j}_{\bmth}} \\
    \approx &\frac{1}{\sigma\sqrt{2 \pi}} \int_{0}^1 d y \sqrt{y} \exp \Big(-\frac{\big(y-\ev{P_j}^2_{\bmth}\big)^2}{2 \sigma^2}\Big) - \abs{\ev{P_j}_{\bmth}},
\end{aligned}
\end{equation}
where $\mathbb{E}^{(m)}(\cdots)$ denotes the expectation value for $m$ shots measurement and $\sigma = \sqrt{(1 - \ev{P_j}^4_{\bmth})/m}$ is the standard deviation of the measurement outcomes.
The standard deviation gets small as $m$ grows so that the bias will vanish.
For a sufficiently large $m$, we further approximate the normal distribution by the uniform distribution across the interval $[\ev{P_j}_{\bmth}^2-\sigma, \ev{P_j}_{\bmth}^2+\sigma]$ to discuss the scaling of the bias.
When $\ev{P_j}_{\bmth}=0$, we have
\begin{equation}
   \begin{aligned} 
B(\wt{\abs{\ev{P_j}_{\bmth}}}) \approx \int_{0}^{\sigma} dy \frac{\sqrt{y}}{2\sigma} = \frac{\sqrt{\sigma}}{3} \sim m^{-1/4}.
   \end{aligned}
\end{equation}
When $\ev{P_j}_{\bmth} \neq 0$ and $m$ is large enough to satisfy $\ev{P_j}_{\bmth}^2 > \sigma$, we have
\begin{equation}
   \begin{aligned} 
B(\wt{\abs{\ev{P_j}_{\bmth}}}) & \approx \int_{\ev{P_j}^2_{\bmth}-\sigma}^{\ev{P_j}^2_{\bmth}+\sigma} dy \frac{\sqrt{y}}{2\sigma} - \abs{\ev{P_j}_{\bmth}} \\
 &\approx -\int_{\ev{P_j}^2_{\bmth}-\sigma}^{\ev{P_j}^2_{\bmth}+\sigma} dy \frac{(y - \ev{P_j}_{\bmth}^2)^2}{16\abs{\ev{P_j}_{\bmth}}^3\sigma}  \\&= -\frac{\sigma^2}{24\abs{\ev{P_j}_{\bmth}}^3} \sim - m^{-1},
   \end{aligned}
\end{equation}
where we used the Taylor expansion of $\sqrt{y}$ at $y=\ev{P_j}_{\bmth}^2 > 0$:
\begin{equation}
    \sqrt{y} \approx \abs{\ev{P_j}_{\bmth}}+\frac{1}{2 \abs{\ev{P_j}_{\bmth}}}(y-\ev{P_j}_{\bmth}^2)-\frac{1}{8 \abs{\ev{P_j}_{\bmth}}^3}(y-\ev{P_j}_{\bmth}^2)^2.
\end{equation}
These equations lead to Eq.~\eqref{eq:bias scaling}.
To mitigate this bias, one prospective approach entails choosing multiple finite values of $m$ and subsequently extrapolating $m$ to $+\infty$, guided by the decay rate of $m^{-1}$ (or $m^{-1/4}$).
We leave it for future investigation.

\begin{table*}[t]
\caption{Geometries and active spaces (if any) of molecules.
$n$ is the number of qubits of the molecular Hamiltonian.
``$(\mr{X}, (x,y,z))$" denotes three dimensional coordinates $x,y,z$ of atom X in units of \AA. \label{tab: geometries}}
 \begin{tabular}{ccl}
 \hline \hline
 Molecule (active space) & $n$ & Geometry  \\ \hline
 \ce{H2} & 4 & (H, (0, 0, 0)), (H, (0, 0, 0.74)) \\
 \ce{H3+} & 6 & (H, (0, 0, 0)), (H, (0, 0, 0.85)), (H, (0, 0.74, 0.43)) \\
 \ce{H4} & 8 & (H, (0, 0, 0)), (H, (0, 0, 1.2)), (H, (0, 0, 2.4)), (H, (0, 0, 3.6)) \\
 \ce{H5+} & 10 & (H, (0, 0, 0)), (H, (0, 0.74, 0.43)), (H, (0, 0.74, -0.43)), (H, (-0.74, 0, 0.43)), (H, (-0.74, 0, -0.43)) \\
 \ce{LiH}(4o,2e) & 8 & (Li, (0, 0, 0)), (H, (0, 0, 1.59)) \\
 \ce{H2O}(4o,4e) & 8 & (O, (0, 0, 0.137)), (H, (0,0.76,0.50)), (H, (0,-0.76,-0.50)) \\
 \ce{NH3}(4o,4e) & 8 & (N, (0,0,0)), (H, (0, 0, 1.01)), (H, (0.95, 0, -0.34)), (H, (-0.48, -0.82, -0.34)) \\
\ce{BeH2}(4o,4e) & 8 & (Be, (0, 0, 0)), (H, (0, 0, 1.33)), (H, (0, 0, -1.33))
\\ \hline \hline
 \end{tabular}
\end{table*}

\section{Details of numerical calculation} \label{appsec:numerics}

\begin{figure}
	\includegraphics[width=8cm]{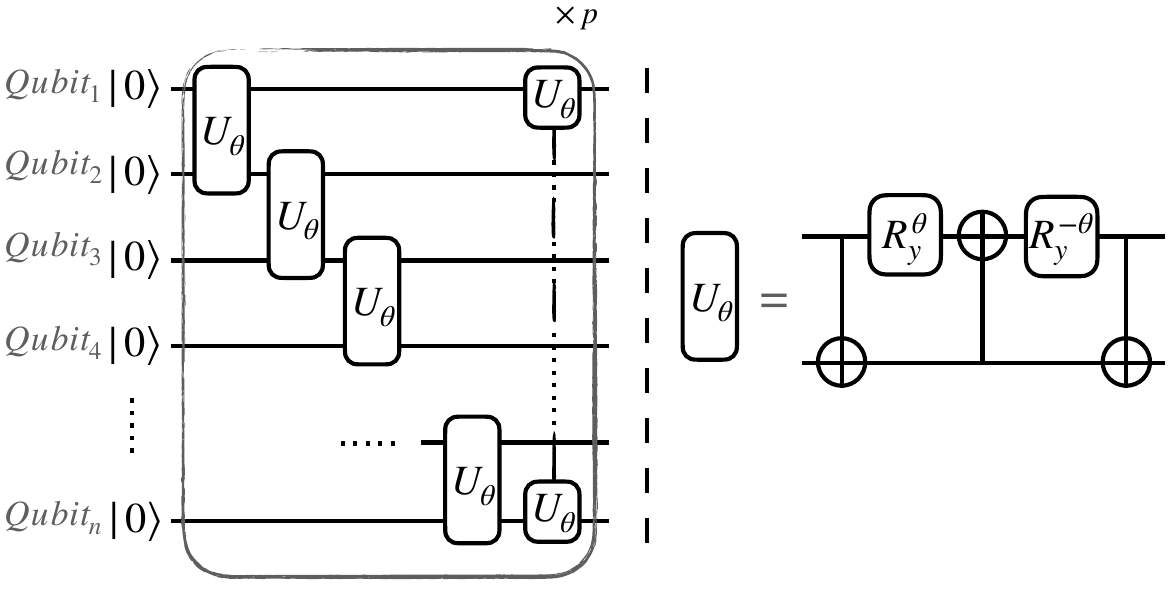}
	\caption{Left: The symmetry preserving real-valued ansatz with $n$ qubits and $p$ layers for constructing the ground state of a molecular Hamiltonian. Right: The $\mr{SO}(4)$ block with three CNOT gates and two $R_y$ rotations with opposite rotation angles.}
	\label{fig:Circuit}
\end{figure}

\subsection{Details of ansatz and molecules}
In the numerical simulations presented in Sec.~\ref{Sec:Results}, we employ the symmetry-preserving real-valued ansatz depicted in Fig.~\ref{fig:Circuit}.
We use molecules with geometries and active spaces listed in Table~\ref{tab: geometries}.
Some of these geometries are chosen as the stable structures at the level of Hartree-Fock/STO-3G, as referenced from the CCCBDB database.

\subsection{Qubit wise commuting grouping}
For the numerical simulations shown in Fig.~\ref{fig:Hn_mol}, we employ the qubit-wise commuting (QWC) grouping~\cite{Kandala2017, Verteletskyi2020} to evaluate the expectation values of the Hamiltonian $H$. 
Consider two $n$-qubit Pauli operators, $Q_1 = \sigma_1^{(1)} \otimes \sigma_1^{(2)} \otimes \dots \otimes \sigma_1^{(n)}$ and $Q_2 = \sigma_2^{(1)} \otimes \sigma_2^{(2)} \otimes \dots \otimes \sigma_2^{(n)}$, where $\sigma_{1,2}^{(j)}$ is a single qubit Pauli operator $I, X, Y, Z$ acting on $j$-th qubit.
We define that $Q_1$ and $Q_2$ are qubit-wise commuting if and only if $\sigma_1^{(j)}\sigma_2^{(j)}=\sigma_2^{(j)}\sigma_1^{(j)}$ for all $j \in \{1,2,\dots,n\}$.
The Pauli operators $\{Q_1, Q_2, \cdots \}$ that are mutually qubit-wise commuting can be simultaneously measured by applying an additional quantum circuit consisting of $O(n)$ one-qubit gates to the state $\ket{\psi}$.
Therefore, we can reduce the number of distinct quantum circuit to evaluate the expectation value of the Hamiltonian $H=\sum_{j=1}^M \lambda_j P_j$ by dividing the Pauli operators $\{P_j \}_{j=1}^M$ into groups of mutually qubit-wise commuting operators.
It should be noted that various grouping methods have been explored in the literature, including those considering usual commutativity and anti-commutativity of the Pauli operators~\cite{mcclean2016, jena2019pauli, Izmaylov2020, zhao2020, Yen2021Cartan}.
The QWC grouping method is adopted in our simulation because it does not require deep and complicated quantum circuits to perform simultaneous measurements of the operators in each group.

The greedy search with sorting the Pauli operators~\cite{crawford2021efficient} is employed when grouping the Pauli operators included in the Hamiltonian. We sort the $M$ Pauli operators $\{P_j \}_{j=1}^M$ in the Hamiltonian by descending order of the absolute values of the coefficients $|\lambda_j|$. The sorted operators are denoted $\{P'_j \}_{j=1}^M$. We assign $P'_1$ to the first group. For $j = 2,3,\dots,k$, if $P'_j$ qubit-wise commutes with all Pauli operators in an existing group, it is assigned to that group. If $P'_j$ does not qubit-wise commute with any Pauli operators in an existing group, a new group is created to house $P'_j$. This procedure is repeated until all Pauli operators are assigned to a group.

\subsection{Choice of $\alpha$ in parameter-shift rule}
Here, we explain the choice of $\alpha=\pi/4$ in the parameter shift rule~\eqref{eq: parameter shift rule} in our numerical calculation.
The parameter shift rule is related to the fact~\cite{Nakanishi2020} that the functional form of the Pauli expectation value with respect to $\theta_l$ is a trigonometric function when other circuit parameters remain constant:
in the simplest cases where the parameter $\theta_l$ in the ansatz $\ket{\psi(\bmth)}$ is an angle of some Pauli rotation gate $e^{-i \frac{\theta_l}{2} Q_l}$ satisfying $Q_l^2=I$, we have
\begin{equation} \label{eq: trig. functional form}
    \ev{P_j}_{\bmth} = a_l \cos \left( \theta_l - \phi_l \right) + c_l, 
\end{equation}
where $a_l, \phi_l$ and $c_l$ are real coefficients depending on $\theta_1, \cdots, \theta_{l-1}, \theta_{l+1},\cdots, \theta_{N_\theta}$.
The parameter shift rule can improve the signal-to-noise ratio when we take larger $\alpha$.
For example, if we use small $\alpha$ like $\alpha = 0.01$, the values $\ev{P_j}_{\bmth + \alpha \delta_l}$ and $ \ev{P_j}_{\bmth - \alpha \delta_l}$ become almost the same so that a lot of measurement shots are required to estimate the gradient ($\propto \ev{P_j}_{\bmth + \alpha \delta_l} - \ev{P_j}_{\bmth - \alpha \delta_l}$) with high accuracy. 
This is why $\alpha=\pi/2$, the largest $\alpha$ considering the periodicity of the function~\eqref{eq: trig. functional form}, is typically used in the literature~\cite{Mitarai2018}.
In our numerical calculation, we observed that some Pauli terms exhibit expectation values approaching $\ev{P_j}_{\bmth} \approx \pm 1$ in the late stages of optimization, or at the vicinity of the exact ground state.
For these Pauli terms, $\abs{\ev{P_j}_{\bmth \pm (\pi/2) \delta_l}}$ sometimes becomes close to zero  (e.g., when $a_l=1$, $c_l=0$ in \eqref{eq: trig. functional form}) and the estimation of it by the joint Bell measurement requires a large number of shots (see Table~\ref{tab:JBM_threshold}). 
Consequently, to strike a balance, we adopt $\alpha = \pi/4$ for our numerical computations presented in Sec.~\ref{Sec:Results}.
In fact, for Pauli terms satisfying $\ev{P_j}_{\bmth} = \pm 1$, $\abs{\ev{P_j}_{\bmth \pm \pi/4 \delta_l}}$ is farther away from zero than $\abs{\ev{P_j}_{\bmth \pm \pi/2 \delta_l}}$ and guaranteed to be larger than $\frac{\sqrt{2}}{2}$,
\begin{equation}
 \abs{\ev{P_j}_{\bmth \pm \pi/2 \delta_l}} = \abs{a_l \cdot \frac{\sqrt{2}}{2} + c_l} \geq \frac{\sqrt{2}}{2},
\end{equation}
because $|a_l| \leq 1$ and $|a_l| + |c_l| \leq 1$ hold due to $\abs{\ev{P_j}_{\bmth}} \leq 1$ for any $\bmth$.

\section{Comparison of JBM-VQE and VQE for the ground-state distribution of $\langle P_j \rangle$}\label{Appendix: Ground-state distribution}
The shot thresholds defined in Section~\ref{Sec:ShotThresholds} and utilized in Section~\ref{Sec:Results} was based on the assumption that $\langle P_j \rangle$ adheres to a uniform distribution within the range of $-1$ and $1$, chosen for its simplicity and relevance to the study.
The numerical results in Section~\ref{Sec:Results} revealed that JBM-VQE significantly accelerates the optimization process compared to the conventional VQE approach. However, it is important to note that in practical scenarios, a variety of factors—such as different ansatzes, initial states, and optimization landscapes—can lead to non-uniform distributions. Consequently, the thresholds for the number of quantum measurement shots may vary. Despite these potential variations, the superiority of our proposed protocol remains evident across any distribution of $\langle P_j \rangle$, particularly when dealing with a larger number of Pauli operators and seeking less precise estimations.

\begin{figure}[b]
	\centering
	\includegraphics[width=0.49\textwidth]{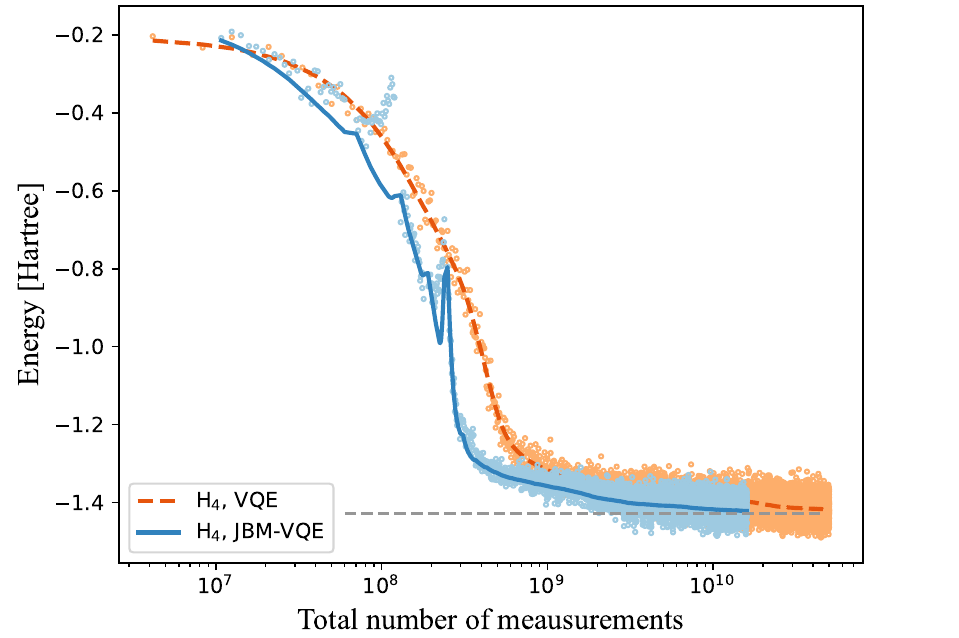}
	\caption{Comparison between the conventional VQE and the JBM-VQE algorithms applied to the molecular Hamiltonian of $\ce{H4}$ for the ground-state distribution of $\langle P_j \rangle$. (a) Total number of measurement shots versus the number of iterations. (b) Optimization curves. The dashed lines and circles denote the actual energy (exact energy expectation value for the parameters at each iteration) and the estimated energy in the conventional VQE, respectively. The solid lines and triangles represent the actual energy and the estimated energy in JBM-VQE, respectively.}  
	\label{fig:H4_mol_gs}
\end{figure}

To illustrate this, we explore a case study involving the $\ce{H4}$ molecule, adopting a more realistic distribution scenario. Here, we consider the actual distribution of the Pauli expectation values for the exact ground state ($|\psi_g\rangle$). These expectation values are represented as $\lambda_1 = \langle\psi_g|P_1|\psi_g\rangle$, $\lambda_2 = \langle\psi_g|P_2|\psi_g\rangle$, and so on, up to $\lambda_{184} = \langle\psi_g|P_{184}|\psi_g\rangle$. For both the conventional VQE and JBM-VQE, the average probabilities of estimating the expectation value within an additive error $\tau_{\text{th}}$ are calculated using the following formulas:
\begin{align}
p^{(\text{av})}(m, \tau_{\text{th}}) = \frac{1}{184}\sum_j p(m, \tau_{\text{th}}, \lambda_j),
\end{align}
and
\begin{align}
q^{(\text{av})}(m, \tau_{\text{th}}) = \frac{1}{184}\sum_j q(m, \tau_{\text{th}}, \lambda_j),
\end{align}
where $p(m, \tau_{\text{th}}, \lambda_j)$ and $q(m, \tau_{\text{th}}, \lambda_j)$ are defined as per Eqs.~\eqref{probability_VQE} and \eqref{probability_JBMVQE}, respectively. For $\tau_{\text{th}} = 0.1$ and $p_{\text{th}} = 0.9$, the shot thresholds, as defined in Definitions ~\ref{Def:Shot_threshold} and ~\ref{Def:JBM_Shot_threshold}, are calculated to be $235$ and $6576$, respectively.
In this experiment, we continue to sample initial circuit parameters from the interval $[0, \pi/5]$ and employ a gradient descent optimization method with a learning rate of $\eta = 0.02$. Figure~\ref{fig:H4_mol_gs} depicts the optimization energy against the number of measurements. The results demonstrate that, for the actual ground-state distribution of $\langle P_j \rangle$, JBM-VQE maintains a noticeable speed advantage over the conventional VQE approach.

\bibliography{ref} 
 
\end{document}